\documentclass[fleqn,twoside]{article}
\usepackage{espcrc2}
\usepackage{graphicx}

\begin{document}

\title{An Algorithm for Obtaining Reliable Priors for Constrained-Curve Fits%
\thanks{Presented by T.\ Draper at Lattice 2003.}%
\thanks{This work is supported in part by the U.S. Department of Energy
        under grant numbers DE-FG05-84ER40154 and DE-FG02-02ER45967.}
}

\author{%
Terrence Draper
  \address[UK]{Department of Physics and Astronomy, 
               University of Kentucky, 
               Lexington, KY 40506, USA},
Shao-Jing Dong
  \addressmark[UK],
Ivan Horv\'{a}th
  \addressmark[UK],
Frank Lee
\address{Center for Nuclear Studies, 
         Dept.\ of Physics, 
         George Washington Univ.,
         Washington, DC 20052, USA}
\address{Jefferson Lab, 
         12000 Jefferson Avenue, 
         Newport News, VA 23606, USA},
Nilmani Mathur
  \addressmark[UK],
Jianbo Zhang
\address{CSSM and Dept.\ of Physics,
         Univ.\ of Adelaide, 
         Adelaide, SA 5005, Australia}
       }

\begin{abstract}
We introduce the ``Sequential Empirical Bayes Method'', an adaptive
constrained-curve fitting procedure for extracting reliable priors.  These are
then used in standard augmented-chi-square fits on separate data.  This better
stabilizes fits to lattice QCD overlap-fermion data at very low quark mass
where {\it a priori\/} values are not otherwise known.  We illustrate the
efficacy of the method with data from overlap fermions, on a quenched
$16^3\times 28$ lattice with spatial size $La=3.2\,{\rm fm}$ and pion mass as
low as $\sim 180\,{\rm MeV}$.
\end{abstract}

\maketitle


\section{Background}

Traditionally, Monte Carlo estimates, $\langle G(t) \rangle$, of two-point
hadronic correlators have been fit to a theoretical model, such as,
\begin{eqnarray*}
      G(t;w_i,m_i) 
& = & 
      \sum_{i=1}^{\infty} w_i e^{-m_i t}
\end{eqnarray*}
by the maximum-likelihood procedure of minimizing the chi-square.

Fitting with a single-source multi-exponential is usually too unstable, so the
default has been the popular single-source single-exponential, wherein one fits
only at $t>t_{\rm min}$ to damp contributions from excited states.  One
compromises between high statistical for large $t_{\rm min}$ and high
systematic errors for small $t_{\rm min}$; lattice alchemy provides various
recipes for making the compromise and estimating the systematic errors.  A
multi-source multi-exponential fit is a much better, albeit more expensive,
alternative; however, the ambiguity of estimating systematic errors through
tuning $t_{\rm min}$ remains.

\vfill

\newpage

``Constrained Curve Fitting''~\cite{Lepage:2001ym,Morningstar:2001je} offers
the alternative of minimizing an augmented chi-square,
\begin{eqnarray*}\label{AugmentedChiSquare}
	\chi_{\rm aug}^2 
= 
	\chi^2 + \chi_{\rm prior}^2 \;\; ; \;\;
	\chi_{\rm prior}^2 
= 
	\sum_i \frac{(\rho_i-\tilde{\rho}_i)^2}
	            {\tilde{\sigma}_i^2}            \label{ChiPrior}
\end{eqnarray*}
where $\rho$ denotes the collective parameters of the fit (e.g.\ $\rho =
\{w,m\}$).  It achieves stability by ``guiding the fit'' with the use of
Bayesian priors, that is, values of the parameters obtained from {\it a
priori\/} estimates $\rho = \tilde{\rho}\pm\tilde{\sigma}$.  It has improved
stability; as data sets are enlarged to include small $t$, many more terms are
added in the fit model until convergence is obtained. The $t_{\rm min}$
systematic error is largely absorbed into the statistical error.  The method
works well if reliable priors are known; nevertheless, stability of the fit
results against choice of prior must be tested, and this reintroduces an
element of subjectivity.

However, with our recent data~\cite{Dong:2003a}, we enter previously unexplored
territory, namely, overlap fermions with exact chiral symmetry at unprecedented
small quark mass and large spatial size.  There is in general no literature
from which to obtain estimates to be used as priors and no reliable model for
estimates of level spacings.

\vfill


\section{Overview}

We propose an adaptive self-contained constrained curve-fitting procedure,
dubbed the ``Sequential Empirical Bayes Method'' method.  We obtain the priors
gradually (from the ground state up) as the data set is monotonically enlarged
(to include earlier and earlier time slices).  The basic ``fixed $\Delta t$''
algorithm is described below for simplicity.  In practice we use a refinement,
the more adaptable ``variable $\Delta t$'' method, results of which are
presented in the next section.
\begin{enumerate}
\item Choose (i) $t_{\rm min}$ and $t_{\rm max}$, the maximum range over which
the fits will be done, and (ii) $t_{\rm start}$, the initial minimum time slice
of the fitting range.
\item Loop on various trial values (``scanning'') around central values $w_1$
and $m_1$ obtained from effective mass.  For each, use an unconstrained fit on
the one-mass-term model (1MTM) to fit over \{$t_{\rm start}$, $t_{\rm max}$\}
to obtain $w_{1}^{(1)}\pm \sigma_{w_1}^{(1)}$ and $m_{1}^{(1)}\pm
\sigma_{m_1}^{(1)}$.  Choose as input for the next step those values which
yield the lowest $\chi^2$.
\item Using these $w_1$, $\sigma_{w_1}$, $m_1$, $\sigma_{m_1}$ as both priors
and initial values, do a constrained-curve fit, using the 1MTM over \{$t_{\rm
start}-1$, $t_{\rm max}$\}, to obtain $w_1^{(2)}$, $\sigma_{w_1}^{(2)}$,
$m_1^{(2)}$, $\sigma_{m_1}^{(2)}$.
\item Using these $w_1$, $\sigma_{w_1}$, $m_1$, $\sigma_{m_1}$ as both priors
and initial values, do a half-constrained fit over \{$t_{\rm start}-2$, $t_{\rm
max}$\} on a 2MTM; the second mass and weight are unconstrained.  Loop on
various trial values for the latter; choose as input for the next step those
values which yield the lowest $\chi^2$.
\item Using these $w_1$, $\sigma_{w_1}$, $m_1$, $\sigma_{m_1}$, $w_2$,
$\sigma_{w_2}$, $m_2$, $\sigma_{m_2}$ as both priors and initial values, do a
fully-constrained fit, using the 2MTM over \{$t_{\rm start}-3$, $t_{\rm
max}$\}, to obtain $w_1^{(4)}$, $\sigma_{w_1}^{(4)}$, $m_1^{(4)}$,
$\sigma_{m_1}^{(4)}$, $w_2^{(4)}$, $\sigma_{w_2}^{(4)}$, $m_2^{(4)}$,
$\sigma_{m_2}^{(4)}$.
\item Repeat the last two steps until all desired mass terms and time slices
are included.  One thus obtains a complete set of priors.
\end{enumerate}


\section{Sample Result}

The $\langle A_4 A_4\rangle$ correlator is dominated by the ground state of the
pseudoscalar channel (pion) over all but the few earliest time slices.  With
the variable $\Delta t$ refinement of the algorithm, rather than deciding {\it
a priori\/} on the number of terms in the fit and adding time slices a fixed
number at a time, one lets the data decide how many time slices to include with
each enlargement of the data by choosing the minimum chi-square over a range of
reasonable possibilities.  Thus since the pion correlator is dominated by the
ground state for many time slices, then many time slices will be automatically
added before an attempt is made to fit the first-excited state.  We find that
the variable $\Delta t$ method works very well for the pion correlators.

\begin{figure}[ht]
  \vspace{-0.5cm}
  \begin{center}
  \includegraphics[angle=0,width=\hsize]{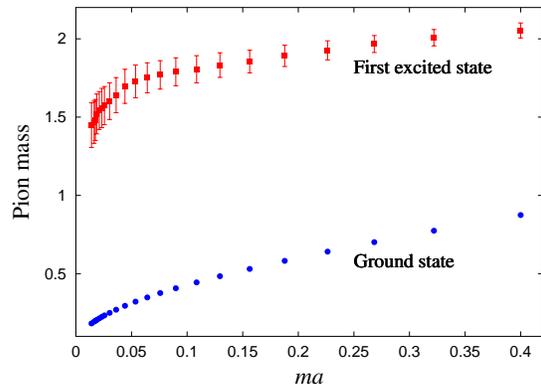}
  \vspace{-1.5cm}
  \caption{Ground and first-excited state pseudoscalar masses as obtained from
   the Sequential Empirical Bayes' Method.}
  \vspace{-1.0cm}
  \end{center}
\end{figure}


\section{Testing the Algorithm}


\subsection{Partitioning the Configurations}

Final results should use the priors on a separate data set, thus preserving the
ideal Bayesian case.  However, empirically this seems to be unnecessary; the
output from the algorithm agrees with the final results.  We have implemented
the following test: We partition the data into two non-intersecting sets of
configurations, $A$ and $B$, with an equal number, $n_{A}=n_{B}=40$ of
configurations in each set.  Using the set $A$ of configurations, we perform
our procedure to obtain priors; we next use this fixed set of priors in the
canonical way to perform a constrained fit separately on data set $B$, on data
set $A$, and on the full set $A\cup B$.  We find no appreciable differences
beyond expected statistical fluctuations.


\subsection{Stability}

The Sequential Empirical Bayes' Method is used to select the priors.  Then a
standard constrained fit with all of the time slices is made.  The number of
terms in the fit model is increased until the fit results converge.

\begin{figure}[ht]
  \vspace{-0.5cm}
  \begin{center}
  \includegraphics[angle=0,width=\hsize]{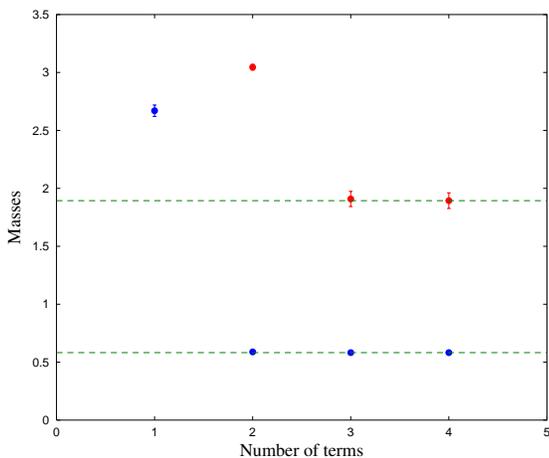}
  \vspace{-1.5cm}
  \caption{Fit values for the lowest two pseudoscalar masses (quark mass
   0.18783), from standard constrained fits (using the priors obtained from the
   Sequential Empirical Bayes' Method).  These are stable as more terms are
   included in the fit model.}
  \vspace{-1.5cm}
  \end{center}
\end{figure}


\subsection{Reconstructing Artificial Data}

The method can successfully reconstruct the parameters of
artificially-constructed data where the true results are known independently of
the fit.

\begin{figure}[ht]
  \begin{center}
  \includegraphics[angle=0,width=\hsize]{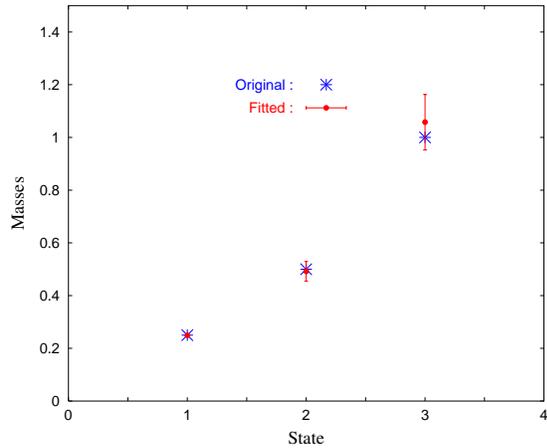}
  \vspace{-1.5cm}
  \caption{Recovery of masses from artificially-constructed data (masses 0.25,
  0.5, 1.0, 1.4, 1.8, and amplitudes 1.20, 1.00, 0.80, 0.60, 0.50).}
  \vspace{-1.0cm}
  \end{center}
\end{figure}

We created a sample of artificial data as a sum of decaying exponentials by
adding an independent Gaussian noise to the function at each value of $t$.  The
fitting procedure was able to reconstruct the means and weights for the ground
and excited states; the actual values were within one measured standard
deviation of the measured means.


\section{Summary}

We have introduced an adaptive self-contained constrained curve-fitting
procedure which produces priors to be used in a standard constrained fit on
different data.  One obtains the priors sequentially as data set is enlarged.
The method's advantages include: it is usable whenever external reliable
estimates of the priors are not available; it can be fully automated; it
reduces human-induced bias; it decreases the frustrating busy-work of fitting;
and it is self-correcting.  We have checked that the method can reconstruct
artificial data, that it is stable against adding more terms, and against
partitioning the data.

For more complete details, see~\cite{Dong:2003b} which provides additional
examples of applicability and outlines further refinements.  The method has
been used successfully to make the first lattice
identification~\cite{Dong:2003a} of the Roper resonance at low quark mass with
exact chiral symmetry.


\end{document}